\documentclass[11pt]{article}
\usepackage{jcapmod}

\usepackage{booktabs}
\usepackage[english]{babel}
\usepackage{amsmath,amssymb,amsbsy,amstext, amsthm, simplewick}
\usepackage{hyperref}
\usepackage{graphicx}
\usepackage{amsfonts}
\usepackage{amssymb}
\usepackage[small]{caption}
\usepackage{upgreek}
 \usepackage{framed}
  \usepackage{xcolor}

\definecolor{Blue}{rgb}{0.25, 0.41, 0.88}
\definecolor{Red}{rgb}{0.92,0.,0.}

\makeatletter
\newlength{\apb@width}
\newcommand{\autoparbox}[2][c]{\settowidth{\apb@width}{#2}\parbox[#1]{\apb@width}{#2}}

\makeatother

\numberwithin{equation}{section}

\def\beq{\begin{equation}}
\def\eeq{\end{equation}}

\def\bea{\begin{eqnarray}}
\def\eea{\end{eqnarray}}

\def\d{{\rm d}}

\def\beq{\begin{equation}}
\def\eeq{\end{equation}}
\def\bea{\begin{eqnarray}}
\def\eea{\end{eqnarray}}

\def\d{{\rm d}}

\def\d{{\rm d}}

\def\r{{\vec r}}
\def\k{{\vec k}}

\def\q{{\vec q}}
\def\p{{\vec p}}
\def\x{{\vec x}}

\def\y{{\vec y}}

\def\Pw{{P^{\rm w}_{\rm in}}}
\def\Pnw{{P^{\rm nw}_{\rm in}}}
\def\Plin{{P^{\rm nw}_{\rm lin}}}

\def\Xw{{\xi^{\rm w}_{\rm in}}}

\def\knl{k_{\rm NL}}

\def\Neff{N_{\rm eff}}

\DeclareRobustCommand{\SkipTocEntry}[4]{}

\begin{document}

\pagenumbering{roman}
\begin{titlepage}
\baselineskip=15.5pt \thispagestyle{empty}

\bigskip\

\vspace{1cm}
\begin{center}
{\fontsize{20.5}{24}\selectfont \sffamily \bfseries Phases of New Physics in the BAO Spectrum}
\end{center}

\vspace{0.2cm}
\begin{center}
{\fontsize{12}{30}\selectfont Daniel Baumann,$^{\bigstar}$ Daniel Green$^{\blacklozenge}$ and Matias Zaldarriaga$^{\clubsuit}$} 
\end{center}

\begin{center}
\vskip8pt
\textsl{$^{\bigstar}$ Institute of Physics, University of Amsterdam, Amsterdam, 1090 GL, The Netherlands}

\vskip8pt
\textsl{$^\blacklozenge$ Department of Physics, University of California, Berkeley, CA 94720, US}

\vskip8pt
\textsl{$^\clubsuit$  Institute for Advanced Study, Princeton, NJ 08540, USA}
\end{center}

\vspace{1.2cm}
\hrule \vspace{0.3cm}
\noindent {\sffamily \bfseries Abstract}\\[0.1cm]
We show that the phase of the spectrum of baryon acoustic oscillations (BAO) is immune to the effects of nonlinear evolution.  This suggests that any new physics that contributes  
 to the initial phase of the BAO spectrum, such as extra light species in the early universe, can be extracted reliably at late times.  We provide three arguments in support of our claim:
First, we point out that a phase shift of the BAO spectrum maps to a characteristic sign change in the real space correlation function and that this feature cannot be generated or modified by nonlinear dynamics.   Second, we confirm this intuition through an explicit computation, valid to all orders in cosmological perturbation theory.
Finally, we provide a nonperturbative argument using general analytic properties of the linear response to the initial oscillations. Our result motivates measuring
 the phase of the BAO spectrum as a robust probe of new physics. 
\vskip10pt
\hrule
\vskip10pt

\vspace{0.6cm}
\newpage
\tableofcontents
\end{titlepage}

\clearpage
\pagenumbering{arabic}
\setcounter{page}{1}

\newpage
\section{Introduction}

Nonlinear evolution is one of the main challenges for using observations of the large-scale structure~(LSS) of the universe as a probe of fundamental physics. This is because {\it i}\hskip 1pt) nonlinear effects are hard to characterize from first principles and {\it ii}\hskip 1pt) they can mimic or distort the signals of interest.
LSS observables that are immune to these nonlinearities are therefore uniquely valuable.  In this paper, we will show that the phase of the spectrum of baryon acoustic oscillations (BAO) is precisely such an observable.  Extracting this phase information from the BAO spectrum would be limited only by statistics and could therefore provide a robust probe of new physics, complementary to the observations of the cosmic microwave background (CMB).

\vskip 4pt
To date, most of the interest in the BAO signal~\cite{Eisenstein:1998tu,Meiksin:1998ra,Eisenstein:2003qy,Hu:2003ti,Eisenstein:2005su,Weinberg:2012es} has focused on obtaining cosmological information from the position of the BAO peak and not from its shape (e.g.~\cite{Alam:2016hwk, Beutler:2016ixs, Ross:2016gvb}).
This is because the shape of the BAO is sensitive to nonlinear effects, which are hard to control at the level required for precision cosmology.  However, as we will show, a part of the shape information, namely that associated with the phase of the power spectrum in Fourier space, is protected from the effects of gravitational nonlinearities and therefore does not need to be discarded when constraining cosmological parameters.

\vskip 4pt
We will consider the effects of both UV and IR modes, as well as their interplay. The fact that short-scale nonlinearities on their own cannot change the BAO phase is easy to understand  from the perspective of an N-body simulation.
 Consider running such a simulation in a box of size smaller than the BAO scale.  Because of the finite spacing of momenta in the box, aliasing removes the oscillatory feature in the power spectrum and turns it into a broadband effect 
as far as modes inside the box are concerned.  Similarly, the nonlinear dynamics of the small-scale modes are smooth in momenta and can, at most, modify the amplitude of the BAO spectrum. 
In fact, the same argument has been applied to establish the robustness of the BAO frequency~\cite{Eisenstein:2006nj} and to motivate the value of the BAO scale as a cosmological probe~\cite{Weinberg:2012es}.  

\vskip 4pt
This means that modes as large as the BAO scale must be present in order to produce more dramatic alterations of the BAO signal.\footnote{Note that this is not purely a statement about modes in the perturbative regime, since  there can be non-trivial couplings between the long and short modes.}  
  However, we will show that these modes only affect the frequency and not the phase of the oscillations.   
  We will provide three arguments in support of our claim: First, we will give an intuitive explanation in position space for why gravitational evolution is insufficient to modify the phase.  
  In short, the gravitational potential away from a localized overdensity is always smoother than the profile associated to the phase shift.  Second, we will show that a change in the phase cannot be generated to all orders in cosmological perturbation theory. 
    Finally, we will provide a nonperturbative argument, using the analytic properties of the linear response\footnote{The small amplitude of the BAO will allow us to focus on the evolution at linear order in the baryon fraction, but to all orders in the underlying matter fluctuations.} to the acoustic oscillations in an inhomogeneous universe.  In all three cases, locality plays a fundamental role in protecting the phase. 

\begin{figure}[t!]
\begin{center}
\includegraphics[width=0.7\textwidth]{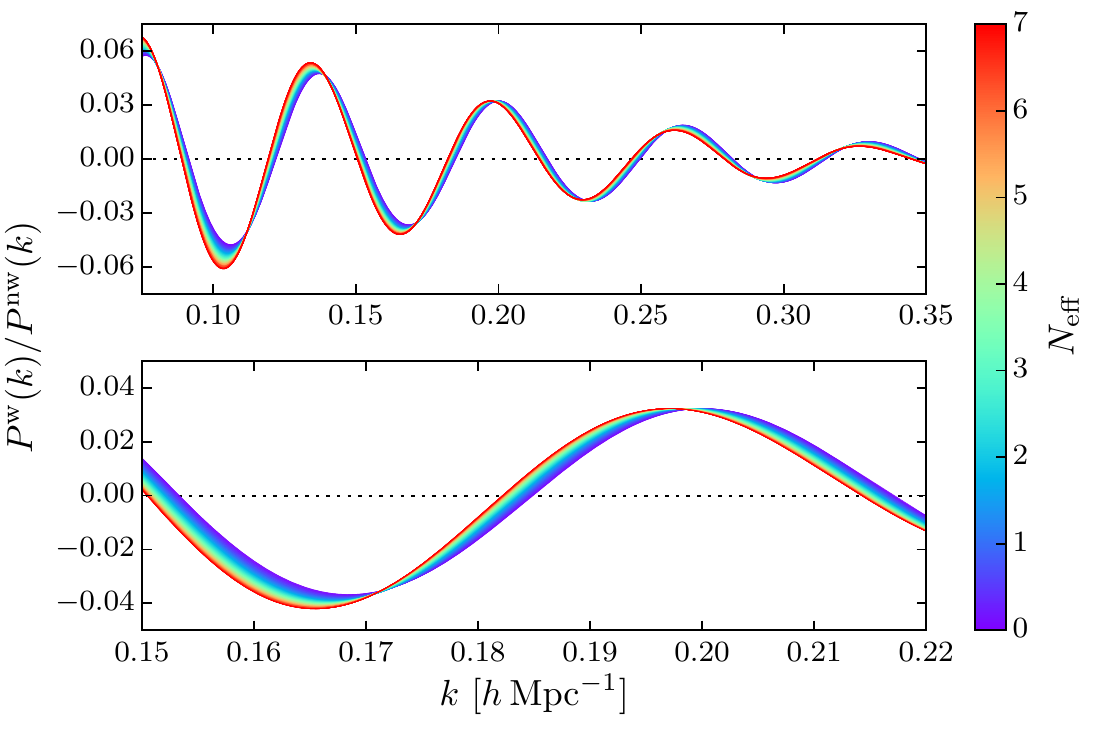}
\caption{Plot of the BAO spectrum $P^{\rm w}(k)$ for varying number of relativistic species $N_{\rm eff}$.   The no-wiggle spectrum $P^{\rm nw}(k)$ has been divided out~\cite{Hamann:2010pw, future}. The photon and baryon densities have been kept fixed, while the dark matter density has been adjusted to keep matter-radiation equality invariant.  The wavenumbers $k$ have been rescaled to remove the effect of $N_{\rm eff}$ on the BAO frequency. The amplitudes of the spectra have been normalized at the peak near $k=0.2\, h\hskip 2pt {\rm Mpc}^{-1}$ which removes the effect of $N_{\rm eff}$ on the amplitude of the oscillations (and some of the effect on the damping envelope). What remains visible is mostly the phase shift of the spectra. }
\label{fig:PhaseShift}
\end{center}
\end{figure}

\vskip 4pt
Our result motivates thinking about new physics that could lead to a phase shift in the acoustic oscillations.
 It is well known that free-streaming relativistic particles, such as neutrinos and other light relics, produce a characteristic phase shift\footnote{For adiabatic initial conditions, the phase of the acoustic oscillations is uniquely fixed. A shift of the phase is therefore a clean signature of non-adiabatic initial conditions or free-streaming relativistic particles~\cite{Bashinsky:2003tk, Baumann:2015rya}.} of the CMB anisotropy spectrum~\cite{Bashinsky:2003tk, Baumann:2015rya}, and that the same phase shift is also imprinted in the BAO spectrum (see Fig.~\ref{fig:PhaseShift}).  
The phase shift due to the neutrinos of the Standard Model has recently been detected in the data of the Planck satellite~\cite{Follin:2015hya, Baumann:2015rya}.  Moreover, future CMB experiments, such as the planned CMB Stage 4 missions~\cite{Abazajian:2016yjj}, will be highly sensitive to the phase of the CMB spectrum\footnote{Future CMB observations will be characterized by more sensitive polarization measurements and improved delensing techniques~\cite{Green:2016cjr}. This will lead to sharper CMB acoustic peaks and an improved sensitivity to the phase of the oscillations.} and have the potential 
to improve current constraints on extra light species by up to an order of magnitude~\cite{Abazajian:2016yjj}.   This corresponds to a percent-level measurement of the radiation density at recombination, which happens to be an interesting threshold: if relativistic species ever were in thermal equilibrium with the Standard Model, their minimal contribution to the radiation density is at the percent level~\cite{Brust:2013xpv, Chacko:2015noa, Baumann:2016wac}.  Reaching this threshold at high significance with CMB observations alone will be challenging~\cite{Abazajian:2016yjj}, so it is encouraging to realize that BAO observations may be an important source of additional information~\cite{future}.  Furthermore, improving measurements of the radiation density has important implications for fundamental physics including the hierarchy problem~\cite{Arkani-Hamed:2016rle,Craig:2016lyx,Chacko:2016hvu}, the strong CP problem~\cite{Ballesteros:2016euj,Baumann:2016wac} and neutrino physics~\cite{Abazajian:2002bj,Hannestad:2012ky,Jacques:2013xr}.

\newpage
\section{Intuition from Position Space}
\label{sec:Real}

Although the phase of the acoustic oscillations is naturally defined in momentum space, much of our physical intuition lives in position space.  In this section,
 we will therefore translate a phase shift in the BAO power spectrum
  to properties of the correlation function in real space.  This will allow us to get an intuitive understanding for why the phase of the spectrum is not altered by nonlinear evolution.   In Section~\ref{sec:Fourier}, we will return to momentum space for a more complete proof of this claim.

\subsection{Preliminaries}

We will take the initial BAO power spectrum to be 
\beq
\Pw(k) = T^{\rm w}(k) \Pnw(k)\, , \quad  T^{\rm w}(k) \equiv A\Big[ \sin(kr_s) \,{\cal D}_\alpha(k) + \beta\,\cos(kr_s) \,{\cal D}_\beta(k) \Big] \, , \label{equ:Pwdef}
\eeq
where $r_s$ is the BAO scale, $A$ is a constant proportional to the baryon fraction $f_b \equiv \rho_b / \rho_m$ and ${\cal D}_{\alpha,\beta}(k)$ are envelope functions\footnote{For simplicity, we will set ${\cal D}_{\alpha,\beta}(k)=1$ in our analytical treatment, although this is not essential.} that encode the damping of the oscillations on small scales.  The superscripts `${\rm w}$' and `${\rm nw}$' stand for ``wiggle" and ``no-wiggle", respectively. The no-wiggle power spectrum,  $\Pnw(k)$, describes the initial conditions for the dark matter in the absence of baryons and the total power spectrum is $P_{\rm in}(k) = \Pw(k) +\Pnw(k)$.  
We will refer to the sine contribution in (\ref{equ:Pwdef}) as the ``neutrinoless BAO feature" and the cosine contribution as the ``phase shift".  The parameter $\beta$ determines the size of the initial phase shift, e.g.~it is proportional to $N_{\rm eff}$ in a theory with extra relativistic species.
We are especially interested in the behavior for $k\to \infty$ where the phase shift is a constant~\cite{Bashinsky:2003tk, Baumann:2015rya}.   The real-space correlation function is 
\begin{align}
\Xw(r) &=  A  \int \frac{k^2 \d k}{2 \pi^2} \, \frac{\sin(kr)}{kr} \, \Big[ \sin(k r_s) + \beta \cos(kr_s) \Big]\,\Pnw(k)\, . \label{equ:Xiw}
\end{align} 
The integration over modes with large momenta will be suppressed due to the rapidly oscillating integrand, unless $r \sim  r_s$, in which case the oscillations cancel between $\sin(k r)$ and $\sin (k r_s)$ or $\cos(k r_s)$.  To describe the limit $k\to \infty$, we are therefore led to consider the behavior of the correlation function near $r = r_s$.  
 
\paragraph{Linear response}

The small baryon fraction, $f_b \ll 1$, will allow us to work perturbatively in the oscillatory part of the initial density contrast $\delta^{\rm w}_{\rm in}$.
 The late-time solution  $\delta^{\rm w}(\x,\tau)$ can therefore be written as a {\it linear response} to $\delta^{\rm w}_{\rm in}$\,:
\beq
\delta^{\rm w}(\x,\tau) =\int \d^3 x' \, G({\color{Red}\x}, {\color{Blue}\x-\x\hskip 1pt'}; \tau) \,\delta_{\rm in}^{\rm w}(\x\hskip 1pt')\, ,\label{equ:14}
\eeq
where $G$ is the response function.\footnote{The first entry of the response function (in {\color{Red} red}) keeps track of the dependence on broken translations, while the second entry (in {\color{Blue} blue}) captures long-range propagation.} Crucially, translation invariance is broken in a given realization of the inhomogeneous universe and therefore $G$ depends on {\color{Red}$\x$} and not just on {\color{Blue}$\x-\x\hskip 1pt'$}. In fact, if the response function $G$ did only depend on $\x-\x\hskip 1pt'$, then in Fourier space it would become a product, $G(\k; \tau)\hskip 1pt \delta^{\rm w}_{\rm in}(\k)$, and it would be trivial to conclude that the evolution does not change the phase of the oscillations.

\paragraph{Power-law universe} 

\begin{figure}[t!]
\begin{center}
\includegraphics[width=0.6\textwidth]{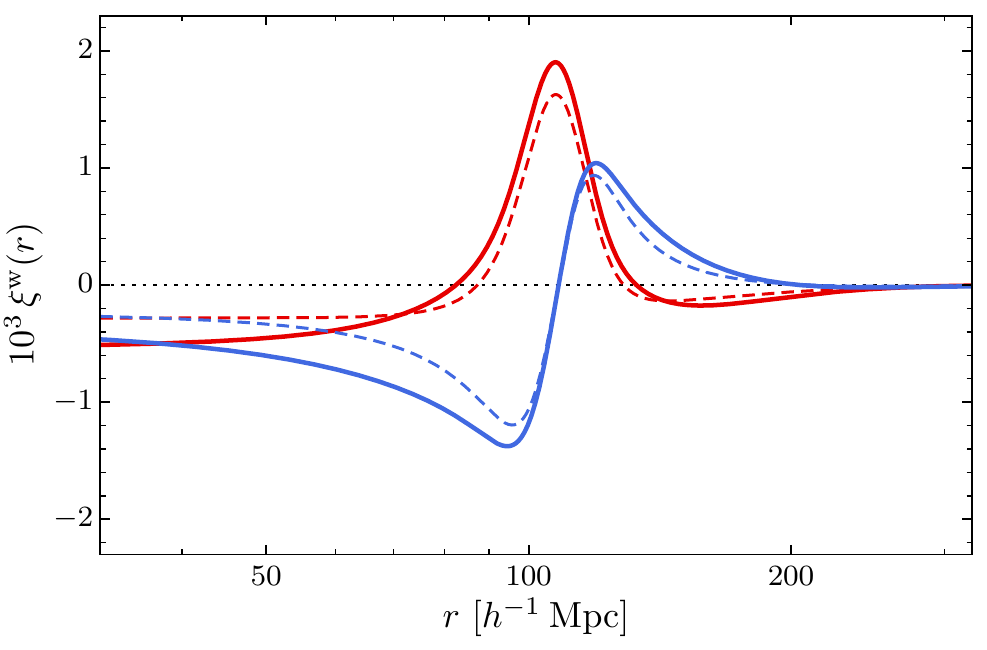}
\caption{Plot of the sine ({\color{Red}red}) and cosine ({\color{Blue}blue}) contributions to the correlation function (\ref{equ:Xiw}) for a toy model with power law index $n=-1.8$ at large $k$.
The model includes the effects of Silk damping and a simple parameterization for the turnover of the power spectrum at low $k$. The dashed lines correspond to the same models with the broadband spectrum taken to be the exact $\Lambda$CDM spectrum. Both sets of curves are more realistic than the pure power-law spectra in (\ref{eq:powerlaw}).  We see qualitatively similar results for the resulting contributions to the correlation function.} 
\label{fig:PS2}
\end{center}
\end{figure}

Let us imagine that the no-wiggle power spectrum locally takes a power-law form
\beq\label{eq:powerlaw}
\Pnw(k) = \frac{k^n}{\knl^{3+n}} \, ,
\eeq
where $\knl$ is a constant momentum scale and $n\approx -1.8$ is a good approximation for the range of scales relevant for the BAO signal.
The resulting wiggle correlation function has a non-trivial dependence on the spectral index $n$.   
Near $r=r_s$, we have
\begin{align}
\Xw(r) &\, \approx\, \left\{ \begin{array}{ll} \displaystyle - \frac{A}{4 \pi^2 r_s \knl^{3+n}} \frac{\Gamma(2+n)}{|r-r_s|^{2+n}} \bigg[ \cos(n \pi/2) + \beta  \sin (n \pi/2) \,{\rm sign}(r-r_s)\bigg]\, , & \quad n<-1 \, ,\\[20pt]
 \displaystyle  +  \frac{A}{4 \pi^2 r_s \knl^{2}} \bigg[ \pi \delta_D(r-r_s) + \beta\, \frac{{\rm sign}(r-r_s)}{|r-r_s|} \bigg]\, ,  & \quad n=-1\, .
\end{array}\right.  \label{equ:Xw}
\end{align}
We see that, for $n\ge -2$, both contributions to $\Xw(r)$ are singular in the limit $r \to r_s$. The distinguishing feature of the phase shift is a sign change at $r=r_s$. 
As shown in Fig.~\ref{fig:PS2}, this property of the phase shift 
 does not depend on the assumption of a power-law universe and holds in more realistic cosmologies.
The choice $n = -1$ is a particularly instructive example because in that case there is no correlation at $r \neq r_s$ for the neutrinoless BAO feature, and only the phase shift contributes. 
Below we will consider both $n = -1$ and $-2 < n < -1$. 

\paragraph{Outline of the argument}  We will first show that purely local evolution in terms of the density contrast $\delta$ cannot generate or modify the phase shift (\S\ref{subsec:local}).
Then, we will ask if this continues to hold when the evolution is nonlocal in $\delta$, but local in terms of the gravitational potential $\Phi = \nabla^{-2} \delta$  (\S\ref{subsec:gravity}). We will see that the change in the correlation due to nonlinear evolution can be written in terms of an effective gravitational potential associated with particles whose distribution is determined by~$\xi^{\rm w}_{\rm in}(r)$.
Finally, we will argue that this potential cannot display the type of singular behavior for $r\to r_s$ that is required for the phase shift.

\subsection{Local Evolution}\label{subsec:local}

Let us first imagine that the evolution is local in $\delta$, i.e.~the response function in (\ref{equ:14}) takes the following form: 
\beq\label{eqn:Glocal}
G({\color{Red}\x}, {\color{Blue}\x-\x\hskip 1pt'}; \tau) =\delta_D({\color{Blue}\x-\x\hskip 1pt'}) \times \sum_m a_m(\tau) \, \big[\delta_{\rm in}^{\rm nw}({\color{Red}\x}\hskip 1pt)\big]^m \, ,
\eeq
and the late-time density contrast is
\beq
\delta^{\rm w}(\x,\tau) = \delta_{\rm in}^{\rm w}(\x\hskip 1pt) \, \sum_m a_m(\tau) \, \big[\delta_{\rm in}^{\rm nw}(\x\hskip 1pt)\big]^m \,.
\eeq
 Working to linear order in the initial BAO feature, 
the correlation function after nonlinear evolution is 
\beq
\xi^{\rm w}(r,\tau) \equiv \langle \delta^{\rm w}(\x,\tau)  \delta^{\rm nw}(\x^{\hskip 1pt\prime},\tau)  \rangle' \,=\, \sum_{l} c_l(\tau) \big[\xi^{\rm nw}_{\rm in}(r)\big]^l \, \Xw(r)  + \kappa(\tau)\, ,
\eeq
where $r \equiv |\x- \x^{\hskip 1pt \prime}|$ and the prime on the correlator denotes dropping the overall delta function.
Factors of $\langle \delta^{\rm nw}_{\rm in}(\x\hskip 1pt) \delta^{\rm nw}_{\rm in}(\x\hskip 1pt) \rangle'$ have been absorbed into $c_l(\tau)$, while $\kappa(\tau)$ includes $\langle \delta^{\rm nw}_{\rm in}(\x\hskip 1pt) \delta^{\rm w}_{\rm in}(\x\hskip 1pt) \rangle'$.  Since $\xi^{\rm nw}_{\rm in}(r)$ is smooth near $r_s$,  we can replace $\sum_l c_l(\tau)\big[\xi^{\rm nw}_{\rm in}(r\sim r_s)\big]^l$ by a constant $\gamma(\tau)$. We then get 
\beq
\xi^{\rm w}(r,\tau) = \gamma(\tau)\,  \Xw(r)  + \kappa(\tau)\, ,
\eeq
which shows that the phase shift is preserved under local evolution.

\vskip 4pt  
Locality also permits terms in the response function~(\ref{eqn:Glocal}) with derivatives acting on $\delta_D(\x-\x\hskip 1pt')$.  Integrating by parts and going through the same steps as before, one finds
\beq
\xi^{\rm w}(r,\tau) = \gamma(\tau)\,  \Xw(r)  + \kappa(\tau) + \omega(\tau) \, R_*\partial_r \Xw(r)  + {\cal O}( R_*^2 \partial^2_r \Xw(r))\, , \label{equ:higher}
\eeq
where $\omega(\tau)$ is some unknown coefficient and $R_*$ is a scale that makes up the dimensions.
The higher-derivative terms in (\ref{equ:higher}) correspond to a perturbative shift in the BAO peak location. For example, the term $\partial_r \Xw(r)$ becomes $k \,\d\Pw/dk$ in Fourier space, which is a frequency shift and not a constant phase shift.  More generally, as long as $R_*$ does not depend strongly on $|r-r_s|$, then the higher-derivative terms in (\ref{equ:higher}) necessarily enter with different powers of $|r-r_s|$ than the terms in (\ref{equ:Xw}) and, as  a result, they don't change the value of $\beta$.  

\subsection{Gravitational Evolution}
\label{subsec:gravity}

One may be concerned that the gravitational potential mediates nonlocal effects when expressed as $\Phi = \nabla^{-2}\delta$. For example, in the case $n = -1$ and $\beta=0$, where $\Xw(r) \propto \delta_D(r-r_s)$, one may wonder if the gravitational potential could turn the delta function into a power law that is visible at $r \neq r_s$, like the phase shift.  However, in order to replicate the phase shift, this power law must also be proportional to ${\rm sign}(r-r_s)$ which is a non-trivial requirement.  

\vskip 4pt
To be concrete, let us write the solution for the late-time density contrast as
\beq
\delta^{\rm w}(\x,\tau)  =  a_1(\tau)\,\Phi_{\rm in}\hskip 1pt \delta_{\rm in} + a_2(\tau)\,\nabla_i \Phi_{\rm in}\hskip 1pt \nabla^i \delta_{\rm in} +a_3(\tau) (\nabla_i \nabla_j \Phi_{\rm in})^2+\cdots\, , 
\eeq
so that the only nonlocality is due to $\Phi_{\rm in}(\x\hskip 1pt)$.  Notice that we are including contributions that explicitly violate the equivalence principle, i.e.~terms proportional to $\Phi_{\rm in}$ rather than just terms built out of the tidal tensor $\nabla_i \nabla_j \Phi_{\rm in}$.  As the inclusion of these terms will illustrate, it is the local evolution in $\Phi$ that ultimately protects the phase and not the equivalence principle.  
To linear order in $\Phi_{\rm in}^{\rm w}$,  the change to the nonlinear correlation function is
  \beq\label{eq:nonlocal}
\Delta \xi^{\rm w}(r,\tau) \approx  c_1(\tau) \, \xi^{\rm nw}_{\rm in}(r) \phi^{\rm w}_{\rm in}(r) + c_2(\tau)\, \partial_r \xi^{\rm nw}_{\rm in}(r) \partial_r \phi^{\rm w}_{\rm in}(r)+ \,\cdots\,,
\eeq
where we have defined \beq
\phi^{\rm w}_{\rm in}(r) \equiv   \langle \Phi^{\rm w}_{\rm in} (\r\hskip1pt) \delta^{\rm nw}_{\rm in}(\vec{0}\hskip 1pt) \rangle' = \int \d^3 r'\, \frac{1}{4\pi}\frac{\Xw(r')}{|\r-\r^{\hskip 2pt \prime}|}  \, .\label{equ:phiw}
\eeq
We notice that $\phi^{\rm w}_{\rm in}(r) $ can be interpreted as an effective gravitational potential sourced by $\xi^{\rm w}_{\rm in}(r')$ [rather than by $\delta^{\rm w}_{\rm in}(r')$].
In fact, this interpretation will provide useful intuition for why the non-locality of gravity does not alter the phase. 

\vskip 4pt
 First, let us consider the case $n=-1$ in~(\ref{equ:Xw}), for which $\Xw(r)\propto \delta_D(r-r_s)$ if $\beta=0$.  The function  $\phi^{\rm w}(r)$ then takes the form of the potential associated with a spherical shell of mass:
\beq
\phi^{\rm w}_{\rm in} \, \propto \, \frac{1}{r} \, , 
\eeq
for $r>r_s$, 
and is constant otherwise.  We see that the limit $r \to r_s$ is smooth and does not display the singular behavior associated to the phase shift.  

\vskip 4pt
Next, we consider the more general case $n< -1$ and/or $\beta \neq 0$.  Because $\Xw(r'=r) \neq 0$, the function $\phi^{\rm w}_{\rm in}(r)$ is no longer of the same form as the potential from an overdensity at $r' = 0$.  However, all of the non-trivial dependence on $r$ beyond that of a point mass must come from integrating~(\ref{equ:phiw}) up to $r' = r$.    Without any further information, we can draw two important conclusions: 
\begin{itemize}
\item[{\it i}\hskip 1pt)] any divergence in $\phi^{\rm w}_{\rm in}(r)$ for $r \to r_s$ will be softer than the divergences in $\Xw(r)$;   
 \item[{\it ii}\hskip 1pt)] any dependence on $r_s$ in $\phi^{\rm w}_{\rm in}(r)$ will be local in $\Xw(r)$.
 \end{itemize}
The first fact follows simply because $\phi^{\rm w}_{\rm in}(r)$ is an integral over~$\Xw(r)$. To establish the second fact, we compare $\phi^{\rm w}_{\rm in}(r)$ and $\phi^{\rm w}_{\rm in}(r+\Delta r)$ and note that any deviation from a $1/r$ potential must be proportional to $\Xw(r')$, with $r\leq r' \leq r+\Delta r$.  Equivalently, we can take derivatives of $\phi^{\rm w}_{\rm in}(r)$ and observe that the most singular terms come from derivatives acting on the upper limit of the integral in~(\ref{equ:phiw}). We conclude that the only reason there can be any dependence on $r_s$ is because $\Xw(r'=r) \neq 0$ and therefore at best (or worst) we would recover the results of \S\ref{subsec:local}. 
  
\vskip 4pt
Finally, let us remark that these results do not depend sensitively on assuming spherical symmetry or on the specific treatment of the long-wavelength modes.  Since we are taking $r \to r_s$, any relatively smooth distribution of particles can locally be treated as an infinite plane.  Since the gravitational potential near an extended plane varies linearly with the distance from the plane, it has no singularity as $r \to r_s$.  The gravitational potential will therefore be smooth near a singular mass distribution, even if the distribution is not perfectly spherical.

\section{Proof in Momentum Space}
\label{sec:Fourier}

In the previous section, we gave an intuitive argument in real space for the protection of the BAO phase.    
While this provided a compelling explanation for our central result, it does not constitute a rigorous proof.  First of all, although the phase shift is unambiguously defined in Fourier space, in real space we required simple power-law universes to cleanly characterize the distinction in the correlation function between the phase shift and the neutrinoless BAO feature.   Second of all, we worked only to linear order in the long-range effects of gravity.  Long-wavelength modes are known to distort the shape of the BAO which then influences the gravitational potential at large distances. In real space, the combined effect of multiple nonlocal interactions is difficult to treat systematically.
Both problems are addressed by going to momentum space, where we have a precise definition of the phase shift and the different physical effects translate into distinct momentum scalings of the response function.

\subsection{Preliminaries}

 In Fourier space, the linear response (\ref{equ:14}) can be written as\hskip 1pt\footnote{As in (\ref{equ:14}), the first entry of the response function (in {\color{Red} red}) captures the effect of broken spatial translations and the second entry (in {\color{Blue} blue}) characterizes long-range propagation.  
 To maintain this distinction after the Fourier transform,  we will sometimes use the colors. }
\beq
\delta^{\rm w}(\k, \tau) = \int \frac{\d^3 q}{(2 \pi)^3}\, G({\color{Red}\k-\q}, {\color{Blue}\q\hskip 2pt}; \tau) \,  \delta_{\rm in}^{\rm w}(\q\hskip 2pt)\,. \label{equ:dw}
\eeq
 Our proof that the phase of this solution is protected will not require the precise form of the response function $G$, but only relies on general analytic properties of $G$.
It is nevertheless instructive to see an explicit expression for the response function in perturbation theory.
A perturbative expansion of the response function reads
\beq
G(\k-\q, \q\hskip 2pt; \tau) \equiv \sum_n G_{(n)}(\k-\q, \q\hskip 2pt; \tau) \, , \label{equ:Gn}
\eeq
where the subscript on $G_{(n)}$ labels the order in perturbation theory.
The time dependence of the $n$-th order solution is approximately given by $G_{(n)}(\k-\q, \q\hskip 2pt; \tau) \approx [D(\tau)]^n\, G_n(\k-\q, \q\hskip 2pt)$, where $D(\tau)$ is the linear growth function. 
The functions $G_n$ are determined in terms of the kernel functions $F_n$ of standard perturbation theory~\cite{Bernardeau:2001qr}:\hskip 1pt\footnote{The functions $G_{n}$ should not be confused with the kernels of the velocity divergence field that are often used in perturbation theory~\cite{Bernardeau:2001qr}.  The only kernels from standard perturbation theory that will appear in this paper are those of the density field, $F_{n}$.}
\begin{align}\label{eq:Pertansatz}
G_n(\k-\q, \q\hskip 2pt) =  \int \frac{\d^3 q_1}{(2\pi)^3} \cdots \frac{\d^3 q_{n-1}}{(2\pi)^3}  &\, n \hskip 1pt F_n(\k-\q\hskip 2pt, \q_1,\ldots, \q_{n-1}) \, \nonumber \\[-4pt]
&\times  \delta^{\rm nw}_{\rm in}(\q_1) \cdots \delta^{\rm nw}_{\rm in}(\q_{n-1})  \,  (2\pi)^3 \, \delta_D\Big(\sum \q_i - \q\hskip 2pt\Big) \, . 
\end{align} 
Up to third order, this can be written as 
\begin{align}
G_1(\k-\q, \q\hskip 2pt)  &\,=\, (2\pi)^3\delta_D(\k-\q\hskip 2pt)\, ,\\[10pt]
G_2(\k-\q, \q\hskip 2pt)  &\,=\, 2\hskip 1ptF_2(\k-\q,\q\hskip 2pt)\, \delta_{\rm in}^{\rm nw}(\k-\q\hskip 2pt)\, , \label{equ:G2}\\[4pt]
G_3(\k-\q, \q\hskip 2pt)  &\,=\, \int \frac{\d^3 q'}{(2\pi)^3} \, 3\hskip 1ptF_3(\k-\q^{\hskip 2pt \prime}-\q, \q^{\hskip 2pt \prime}, \q\hskip 2pt) \, \delta_{\rm in}^{\rm nw}(\k-\q^{\hskip 2pt \prime}-\q\hskip 2pt) \,\delta_{\rm in}^{\rm nw}(\q^{\hskip 2pt \prime})\, . \label{equ:G3}
\end{align}
An explicit expression for $F_2$ will be given below, that for $F_3$ can be found in~\cite{Goroff:1986ep} and recursion relations for the higher-order $F_n$'s are presented in~\cite{Bernardeau:2001qr}.

\vskip 4pt
We write the oscillatory part of the initial conditions as
\beq
 \delta_{\rm in}^{\rm w}(\q\hskip 2pt) = A(q) \sin(q r_s+\varphi_{\rm in}) \,\delta^{\rm nw}_{\rm in}(\q\hskip 2pt)\, ,
 \eeq
 where the envelope $A(q)$ describes the damping of oscillations on small scales and $\varphi_{\rm in}$ is the initial phase. For simplicity, we will set
$A(q)\to A = const.$ and $\varphi_{\rm in}=0$, although this is not essential to our argument.  The wiggle  spectrum can then be written as  
 \begin{align}
P^{\rm w}(k,\tau) 
&=  A\hskip 1pt \Plin(k,\tau) \int \frac{\d^3 q}{(2 \pi)^3}\, g({\color{Red}\k-\q},{\color{Blue}\q}\hskip 2pt; \tau)\,  \sin(q r_s)  \,+\,\{\k\to -\k\hskip 1pt\}\, ,  \label{equ:Pwres}
\end{align}
where $\Plin(k,\tau) \equiv D^2(\tau) P^{\rm nw}_{\rm in}(k)$ is the linearly-evolved no-wiggle spectrum and the kernel $g(\k-\q,\q\hskip 2pt;\tau)$  is defined as
\beq
g(\k-\q,\q\hskip 2pt;\tau) \equiv \frac{\langle   G(\k-\q, \q\hskip 2pt; \tau) \,  \delta^{\rm nw}_{\rm in}(\q\hskip 2pt)\, \delta^{\rm nw}(-\k\hskip 1pt, \tau)  \rangle'}{\Plin(k,\tau)}\, . \label{equ:gdef}
\eeq
Crucially, the kernel $g(\k-\q,\q\hskip 2pt; \tau)$ is a smooth function at large momenta and does not depend on the scale $r_s$.  Moreover, we will be insensitive to the behavior of the kernel function at low momenta, including the delta function for vanishing momenta that is required when the response is translation invariant.  

\begin{figure}[t!]
\begin{center}
\includegraphics[width=0.6\textwidth]{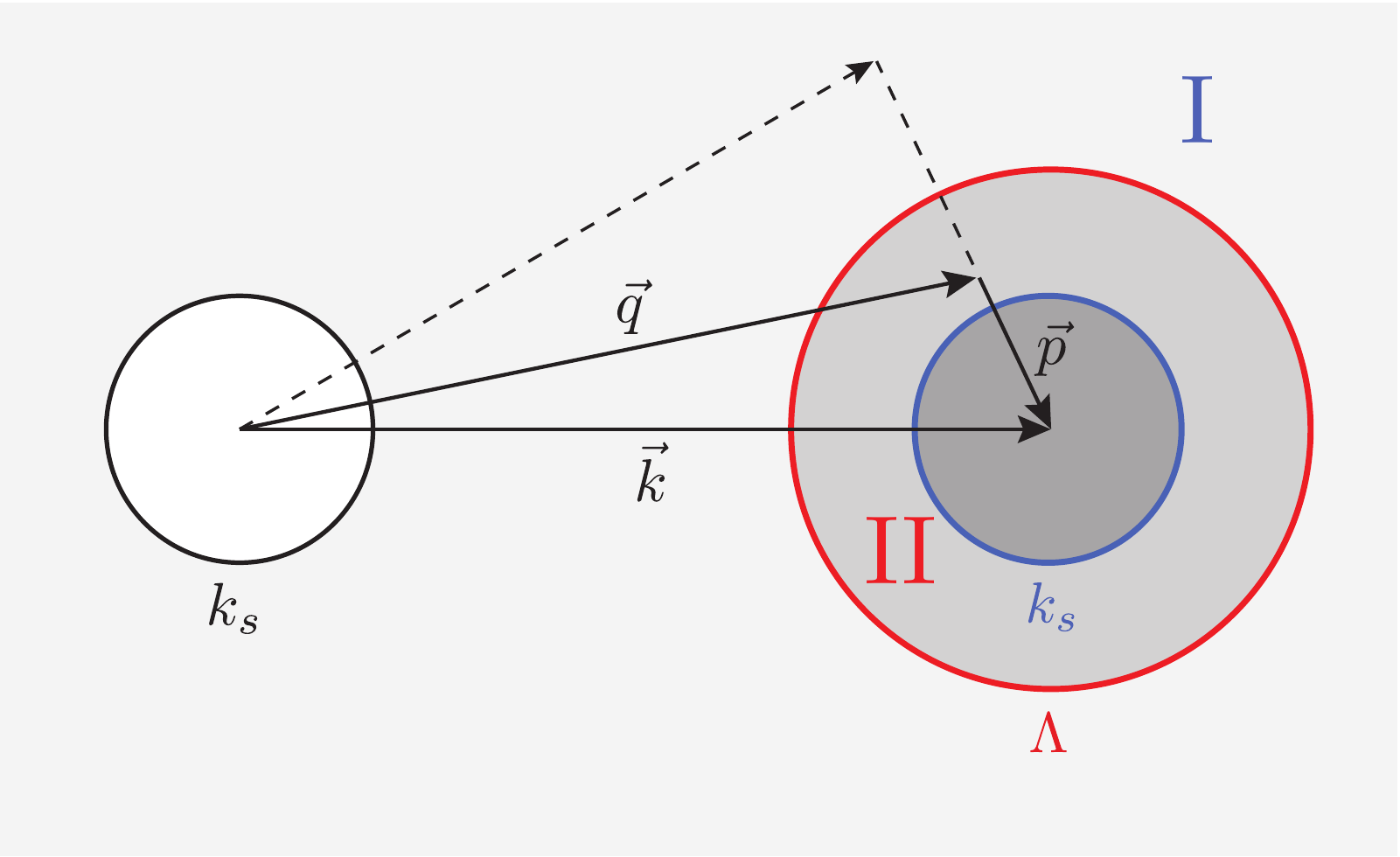}
\caption{Illustration of the domain of integration of $\q$ for fixed $\k$. We divide the $\q$-plane into two regions: I) $p \equiv |\k-\q\hskip 2pt| > k_s$ (outside the {\color{Blue}blue} circle) and II) $p< \Lambda < k$ (inside the {\color{Red}red} circle). These regimes are considered in \S\ref{sec:short} and \S\ref{sec:long}, respectively. In the overlap region, $k_s< p < \Lambda$, both treatments apply.} 
\label{fig:qplane}
\end{center}
\end{figure}

\vskip 4pt
We are interested in the evolution of the wiggle spectrum (\ref{equ:Pwres}) for $k > k_s \equiv r_s^{-1}$.  Nonlinearities induce couplings to modes with $\q\ne \k$.  
 Only fluctuations with $q > k_s$ can lead to oscillations in~$k$, so we will focus on those.   
We will treat separately the effects of short modes, $|\k-\q\hskip 2pt| > k_s$, and long modes, $|\k-\q\hskip 2pt| < \Lambda < k$, where $\Lambda$ is an arbitrary scale separating UV and IR modes.
Figure~\ref{fig:qplane} illustrates the different regions of interest in the $\q$-plane.

\paragraph{Outline of the argument} Our argument proceeds in multiple steps.  First, we will show~(\S\ref{sec:short}) that short-wavelength modes alone do not affect either the frequency or the phase of the BAO spectrum.  This formalizes the intuition that the BAO feature appears at large distances and is therefore immune to short-distance effects. Second, we will prove (\S\ref{sec:long}) that long-wavelength modes do not change the phase, although they modify the frequency. We will treat the long modes first in perturbation theory~(\S\ref{sec:PT}) and then nonperturbatively (\S\ref{sec:NP}).  
Since long-wavelength modes are known to produce a frequency shift (e.g.~\cite{Crocce:2007dt,Nishimichi:2008ry,Padmanabhan:2009yr,Mehta:2011xf,Sherwin:2012nh,Baldauf:2015xfa, Blas:2016sfa}), or equivalently a $k$-dependent phase shift, it is not obvious a priori that they could not also lead to a constant phase shift. 
Fortunately, at every step, we will find that only odd powers of $k$ appear in the phase shift and therefore a constant shift is not possible.  This important odd/even counting of the $k$-dependence is a consequence of {\it locality} ({analyticity} in $\k$\hskip 2pt) and {\it rotational invariance}.   
Implementing locality will be complicated by the long-range nature of the gravitational influence.

\subsection{Short Modes}
\label{sec:short}

First, we consider the effects of small-scale modes. Specifically, we are interested in the behavior when all relevant momenta are large, i.e.~$q \gg k_s$ and $p \equiv |\k -\q\hskip 2pt| \gg k_s$. This corresponds to region~I in Fig.~\ref{fig:qplane}.  
 In order to diagnose whether short-scale nonlinearities can affect  the oscillations,  we perform a
shift $\k \to \k + k_s \hskip 1pt \vec{\alpha} $, with $|\vec{\alpha}\hskip 1pt| = {\cal O}(1)$.  
If the result of (\ref{equ:Pwres}) were an oscillating function of $kr_s$, then the shift in $k$ would lead to an order-one change in the answer.  However, since $k, |\k-\q\hskip 2pt| \gg k_s$, we have\hskip 1pt\footnote{From now on, functions without an explicit time argument are to be evaluated at an arbitrary time $\tau$, except for those with a subscript `in' which are defined at a fixed initial time $\tau_{\rm in}$.}
\begin{align}
g\big(\k-\q,\q\hskip 2pt\big) &\ \to\ g\big( (\k -\q\hskip 2pt) + k_s \hskip 1pt \vec \alpha , \q \hskip 2pt\big) \nonumber\\[4pt]
&\ =\ g\big(\k-\q,\q\hskip 2pt\big)  + k_s \hskip 1pt \vec \alpha   \cdot  \partial_{\k} \hskip 2pt g\big(\k-\q,\q\hskip 2pt\big)+ \cdots \nonumber\\[4pt]
&\ \approx \ g\big(\k-\q,\q\hskip 2pt\big) + {\cal O}((k/k_s)^{-1})  \, \frac{\vec \alpha \cdot \k}{k} \, g\big(\k-\q,\q\hskip 2pt\big)+\cdots  \, ,
\end{align}
where we have used that $g(\k-\q,\q\hskip 2pt)$ is a smooth function of its arguments at high momenta, so that $\partial_\k \sim \k /k^2$. Performing the integral\footnote{When $|\k -\q \hskip 2pt| \ll k$, there may be enhanced contributions where $\k /k^2 \to (\k -\q\hskip2pt)/|\k -\q \hskip 2pt|^2$.  These special values of $\q$ may or may not contribute significantly to the integral, depending on the form of $g(\k-\q,\q\hskip 2pt)$.  Nevertheless, since we are considering $|\k-\q \hskip 2pt| \gg k_s$, the higher-order terms in the Taylor expansion will always be suppressed.} in~(\ref{equ:Pwres}), the result of the shift in the momentum is found to be small
\beq
\frac{\big|P^{\rm w}(|\hskip 1pt\k + k_s \hskip 1pt \vec\alpha\hskip 1pt|) - P^{\rm w}(k)\big|}{P^{\rm nw}(k)} \ \lesssim\ {\cal O} \left({k_s/k}\right) \ll 1\,  .
\eeq
This shows that the mode coupling with short-scale fluctuations only produces broadband effects and does {\it not} change the frequency or phase of the oscillations.

\vskip 4pt
The robustness of the BAO frequency to the effects of short modes
 was famously argued in~\cite{Eisenstein:2006nj}.  The basic intuition is that, although the oscillations appear at high momenta, they only affect the correlation function at distances of order $r_s \sim$ 150~Mpc, which is a much larger than the scale associated with nonlinearities, $r_{\rm NL} \sim 10$~Mpc. This can also be seen by running an N-body simulation in a box of size smaller than the BAO scale, $L \ll r_s$. Imposing periodic boundary conditions, the allowed momenta inside the box satisfy $\Delta k = 2\pi/L \gg {2\pi/r_s}$, i.e.~they are spaced by more than the 
period of the oscillatory feature.  As a result,  the evolution inside the box cannot tell the difference between a high-frequency oscillation and a smooth power spectrum and therefore only the overall normalization of the oscillations can be affected by the nonlinear evolution of the short modes.

\subsection{Long Modes}
\label{sec:long}

Next, we consider the effects of long-wavelength modes, with $|\k-\q\hskip 2pt| \lesssim \Lambda < k$. This corresponds to~region~II in Fig.~\ref{fig:qplane}.  To analyse this regime, it is useful to change the integration variable to $\p \equiv \k-\q$, so that 
\beq\label{eq:Pwlow}
P^{\rm w}(k) = A\hskip 1pt \Plin(k) \int \frac{\d^3 p}{(2\pi)^3}\,  g({\color{Red}\p},{\color{Blue}\k-\p}\hskip 2pt)\,  \sin(|\k-\p\hskip 2pt| r_s ) \,+\, \{ \k \to -\k \hskip 1pt\} \, ,  
\eeq
where the region of interest is  now $p < \Lambda < k$.  It is also convenient to write
\beq
\sin(|\k-\p\hskip 2pt|r_s) = \sin(kr_s)  \cos\big((|\k-\p\hskip 2pt| - k)r_s\big) - \cos(kr_s) \sin\big((|\k-\p\hskip 2pt| - k)r_s\big)\,  ,
\label{equ:sin-exp}
\eeq
so that~(\ref{eq:Pwlow}) takes the form 
\begin{align}
\boxed{P^{\rm w}(k)\,=\,  A\hskip 1pt \Plin(k)\left[ f(k) \cos(k r_s) + \tilde f(k) \sin(k r_s) \right]}\  , \label{eq:Pwlow2}
\end{align}
where 
\begin{align} 
f(k) &\equiv - \int \frac{\d^3 p}{(2\pi)^3} \, g(\p,\k-\p\hskip 2pt)\, \sin\big((|\k-\p\hskip 2pt| - k)r_s\big) \,+\, \{ \k \to -\k \hskip 1pt\} \, , \label{equ:ff}\\ 
\tilde f(k) &\equiv + \int \frac{\d^3 p}{(2\pi)^3} \, g(\p,\k-\p\hskip 2pt)\, \cos\big((|\k-\p\hskip 2pt| - k)r_s\big) \,+\, \{ \k \to -\k \hskip 1pt\} \, . \label{equ:tff}
\end{align}
Since the changes to the BAO spectrum from nonlinear evolution are known to be small, we can treat $f(k)$ as a small parameter.  The wiggle spectrum (\ref{eq:Pwlow2}) can then be written as
\beq
\boxed{P^{\rm w}(k)\,\approx\,  A\hskip 1pt \Plin(k)  \tilde f(k) \sin(k r_s + \varphi(k)) }\ , \quad {\rm where} \quad \varphi(k) \equiv f(k) / \tilde f(k)\ll 1\,.
\eeq
It is useful to note that any $k$-dependence that is common to both $f(k)$ and $\tilde f(k)$ does not alter the phase $\varphi$. To reduce clutter, we will sometimes drop these common factors.
We will write the integrands in (\ref{equ:ff}) and (\ref{equ:tff}) as an expansion in powers of $p/k$, keeping the leading order terms for both odd and even powers of $\mu \equiv  \k\cdot \p\hskip 2pt/(kp)$ separately.  The odd and even terms behave differently under the integration over $\mu$ and therefore the relative  $p/k$ suppression can be offset by the angular integration.   This leads to
\begin{align}
f(k) &\approx   \int \frac{\d^3 p}{(2\pi)^3} \left[\, g^-(\p,\k\hskip 1pt) \, \sin( \mu p r_s)-  g^+(\p,\k\hskip 1pt)  \,pr_s \cos(\mu p r_s) \, \frac{1-\mu^2}{2}\, \frac{p}{k}\,   \right]  , \label{equ:ff2} \\
\tilde f(k) &\approx  \int \frac{\d^3 p}{(2\pi)^3}\left[ \, g^+(\p,\k\hskip 1pt) \, \cos( \mu p r_s)+  g^-(\p,\k\hskip 1pt)  \,  p r_s \sin(\mu p r_s) \,\frac{1-\mu^2}{2}\, \frac{p}{k}  \, \right]  ,
\end{align}
where we have defined $g^\pm(\p,\k\hskip 1pt) \equiv g(\p,\k-\p\hskip 2pt) \pm g(\p,-\k-\p\hskip 2pt)$.  
We will show that $f(k)$ and $\tilde f(k)$ only contain odd and even powers of $k$, respectively, which excludes the possibility of a constant phase shift.
We will first demonstrate this in perturbation theory and then present a nonperturbative argument.

\subsubsection{Perturbative Argument}
\label{sec:PT} 

In perturbation theory, the density contrast  can be written as
\beq\label{eq:SPT}
\delta(\k,\tau) = \sum_n \delta_{(n)}(\k,\tau) \, ,
\eeq
where $\delta_{(n)}(\k,\tau) \approx [D(\tau)]^n \,\delta_n(\k\hskip 1pt)$ and the $n$-th order solution $\delta_n$ is a convolution of $n$ powers of the initial density contrast $\delta_{\rm in}$\hskip 2pt:
\beq
\delta_n(\k\hskip 1pt) \,=\,  \int \frac{\d^3 p_1}{(2\pi)^3} \cdots\frac{\d^3 p_n}{(2\pi)^3}  \, F_n(\p_1,\ldots, \p_n) \,  \delta_{\rm in}(\p_1) \cdots \delta_{\rm in}(\p_n) \,  (2\pi)^3 \, \delta_D\Big(\sum \p_i - \k\hskip 2pt\Big) \, . \label{eq:Fn}
\eeq
The kernel function of the second-order solution is
\begin{align}
F_2(\p,\k-\p\hskip 2pt) &\equiv \frac{5}{7} \,+\,\frac{1}{2} \left(\frac{1}{p^2} \,+\, \frac{1}{|\k-\p\hskip 2pt|^2}\right)  \p \cdot (\k-\p\hskip 2pt) + \frac{2}{7} \frac{[\p\cdot (\k-\p\hskip 2pt)]^2}{p^2|\k-\p\hskip 2pt|^2}\
\\[4pt]
&= \frac{k}{p} \left[\frac{\mu}{2} + \frac{3+4\mu^2}{14} \frac{p}{k} - \frac{15\mu-8\mu^3}{14} \left(\frac{p}{k}\right)^2 + \cdots \right] ,
\end{align}
where the second line is an expansion in the small ratio $p/k$. 
Notice that the coefficients of the even (odd) powers of $p/k$ are even (odd) functions of $\mu$.  This property is essential for our argument and continues to hold for the kernels $F_n$ of the $n$-th order solutions $\delta_n$~\cite{Bernardeau:2001qr}.

\paragraph{One-loop order} 
The oscillatory part of the second-order solution, $\delta_2^{\rm w}$, takes the form (\ref{equ:dw}), with the response function given by~(\ref{equ:G2}). Correlating this with $\delta_2^{\rm nw}$, we find 
\begin{align}
g_{22}(\p, \k-\p\hskip 2pt) &\equiv  \frac{\langle   G_{(2)}(\p, \k-\p\hskip 2pt) \,  \delta^{\rm nw}_{\rm in}(\k-\p\hskip 2pt)\, \delta^{\rm nw}_{(2)}(-\k\hskip 1pt)  \rangle'}{\Plin(k)} \nonumber \\
&= 2 \hskip 1pt \Plin(p)\, F_2^2(\p,\k-\p\hskip 2pt)\,  \frac{\Plin(|\k-\p\hskip 2pt|)}{\Plin(k)}   \nonumber \\[4pt]
&= 2\hskip 1pt \Plin(p)\,\frac{k^2}{p^2}\left(\frac{\mu^2}{4}+ \left[\frac{3\mu+4\mu^3}{14} - \frac{\mu^3}{4} \frac{d \ln \Plin}{d\ln k}\right]\frac{p}{k}+\cdots\right) . \label{eq:g22}
\end{align}
Substituting this into the integrand of (\ref{equ:ff2}), we get
\begin{align}
& \left[\, g_{22}^-(\p,\k\hskip 1pt) \, \sin( \mu p r_s)-  g_{22}^+(\p,\k\hskip 1pt)  \,p r_s \cos(\mu p r_s) \,  \frac{1-\mu^2}{2}\, \frac{p}{k}\,   \right]  \\[3pt]
&\ \ = 2\hskip 1pt \Plin(p)\, kr_s \underbrace{\left(  \left[- \frac{3\mu+4\mu^3}{7} + \frac{\mu^3}{2} \frac{d \ln \Plin}{d\ln k}\right] \frac{\sin(\mu p r_s)}{pr_s} + \frac{\mu^4-\mu^2}{4} \cos(\mu pr_s)\right)}_{\displaystyle \equiv {\cal I}(\mu,p,k)} +\, {\cal O}((kr_s)^{-1})\,. \nonumber
\end{align}
Ignoring the weak $k$-dependence of ${\cal I}(\mu,p,k)$, we then find
\beq
\begin{aligned}
f_{22}(k) &\,\approx\, kr_s \int_0^\Lambda \frac{\d p}{2\pi^2} \, p^2 \Plin(p)\int_{-1}^1 \d \mu\,\hskip 2pt {\cal I}(\mu,p) \ +\ {\cal O}((kr_s)^{-1}) \\[6pt]
&\, \sim\, kr_s\, \sigma^2\ +\ {\cal O}((kr_s)^{-1})\, , \label{equ:f22}
\end{aligned}
\eeq
where we have defined the variance of the IR fluctuations
\beq
\sigma^2(\Lambda) \equiv  \int_0^\Lambda \frac{\d p}{2\pi^2}\, p^2 \Plin(p)\, .
\eeq
The key property of the solution (\ref{equ:f22}) is the absence of a $k^0$ term. The leading term is proportional to $k$ and corresponds to a frequency shift, $r_s \to r_s(1+\sigma^2)$, rather than a phase shift.

\vskip 4pt
At one-loop order, we should also consider the correlation between $\delta_3^{\rm w}$ and $\delta_1^{\rm nw}$.  Using the third-order response function (\ref{equ:G3}), we find
\begin{align}
g_{31}(\p, \k-\p\hskip 2pt)  & \equiv  \frac{\langle   G_{(3)}(\p, \k-\p\hskip 2pt) \,  \delta^{\rm nw}_{\rm in}(\k-\p\hskip 2pt)\, \delta^{\rm nw}_{(1)}(-\k\hskip 1pt)  \rangle'}{\Plin(k)}  \nonumber \\[4pt]
&=  \, (2\pi)^3 \delta_D(\p\hskip 2pt) \int \frac{\d^3 p'}{(2\pi)^3}\, 3 \hskip 1pt F_3(\k , \p^{\hskip 2pt\prime} , -\p^{\hskip 2pt\prime}\hskip 2pt)\, \Plin(p') \, ,
\end{align}
where the delta function in $\p$ arises from a self-contraction in $\delta_3^{\rm w}$.  Substituting this into (\ref{equ:ff}) and (\ref{equ:tff}), we find $f_{31}(k) =0$ and $\tilde f_{31}(k) \ne0$, respectively. This shows that the third-order solution only leads to a change in the amplitude of the oscillations,  but not the frequency or phase.  The same is true for $f_{13}(k)$ because $G_{(1)}(\p, \k-\p\hskip 2pt) \propto \delta_D(\p-\k\hskip 1pt)$ and $P^{\rm w}_{\rm lin}(k)$ trivially factors out.

\paragraph{All-orders extension} 
 It is relatively straightforward to extend this perturbative argument to all orders (see also~\cite{Sugiyama:2013gza,Vlah:2015zda} for related discussion). 
The oscillatory part of the $n$-th order solution, $\delta_n^{\rm w}$, takes the form (\ref{equ:dw}), with the response function given by~(\ref{eq:Pertansatz}). 
Correlating this with $\delta_m^{\rm nw}$, we have 
\begin{align}
g_{nm}(\p, \k-\p\hskip 2pt) &\equiv \frac{\langle   G_{(n)}(\p, \k-\p\hskip 2pt) \,  \delta^{\rm nw}_{\rm in}(\k-\p\hskip 2pt)\, \delta^{\rm nw}_{(m)}(-\k\hskip 1pt)  \rangle'}{\Plin(k)} \, .
\end{align}
Note that we require $n+m$ to be even in order to contract all factors of $\delta_{\rm in}$, otherwise $g_{nm} =0$.  
For $m=1$, we get $g_{n1} \propto \delta_D(\p\hskip 2pt)$, for any $n$, which
only contributes  to a change in the amplitude of the oscillations.  For $m \geq n \geq 2$, we instead have 
\begin{align}
g_{nm}(\p, \k-\p\hskip 2pt) &\,= \,\sum_{j}\prod_{i=1}^{l}\int \frac{\d^3 p_i}{(2\pi)^3}\
\kappa_{j} \, F_n(\{\p_a \}, \{ \p_{b}, -\p_{b}\}, \k-\p\hskip 2pt) F_m(\{-\p_{a} \},\{ \p_{c}, -\p_{c}\}, -\k+\p\hskip 2pt)\nonumber \\[-10pt]
&\hspace{3.5cm}\times \,\Plin(p_1) \cdots \Plin(p_l) \,\frac{\Plin(|\k-\p\hskip 2pt|)}{\Plin(k)}  \, , \label{eq:gnm0}
\end{align}
where the sum  runs over $j=0,2,\ldots, n-1$ or $j=1,3,\ldots, n-1$, when $n$ is odd or even, respectively, and $\kappa_{j}$ are combinatorial factors whose explicit form will not be important. 
The subscripts on the momentum entries of $F_n$ and $F_m$ run over
$a =1,\ldots,j$, $b=j+1, j+3,\ldots,n-2$, and $c =j+1, j+3,\ldots, m-2$.  We have integrated over $l \equiv (n-1+m)/2$ internal momenta $\p_i$, with $\sum_{a=1}^{j} \p_a = \p$ (the sum over the remaining momenta vanishes by definition).  
  For $m < n$, the result is the same as~(\ref{eq:gnm0}) if we exchange $n \leftrightarrow m$ and let $\k, \p \to -\k , -\p$.  The main difference is only the number of self-contractions inside $F_n$ and $F_m$ which has no impact on the $k$-dependence.  Since the resulting behavior of $f_{nm}(k)$ will be identical in both cases, we will restrict to $m \geq n$.  

\vskip 4pt
Recall that we are working in the limit $p \ll k$.  A priori, this does not require that  $p_i \ll k$, since we could have $\p_i \approx -\p_{i+1}$ and $p_i \gg p$, so that $p = |\sum_i \p_i | \ll \sum_i |\p_i|$.  However, 
in this limit,
$F_{n} F_{m} \propto p_i^{-4}$~\cite{Goroff:1986ep} and we therefore do not get a significant contribution to the integral.  We can therefore focus on the regime where $p_i \ll k$, for all $i$.

\vskip 4pt
To find the result for a general term in  (\ref{eq:gnm0}), we perform a Taylor expansion in $p_i/k \ll 1$:
\begin{align}
g_{nm}(\p, \k-\p\hskip 2pt) 
&\,\supset\, \Bigg( \prod_{i=1}^{l}\int \frac{\d^3 p_i}{(2\pi)^3} \, (\k \cdot \p_i\hskip 2pt )^{s_{i}} \Bigg) k^{2 t} \, \frac{\Plin(|\k-\p\hskip 2pt|)}{\Plin(k)}  \,{\cal F}_{nm}(\p_1,\ldots, \p_l) \, , \label{equ:gnm}
\end{align}
where $t$ is an integer (positive or negative), $s_{i}$ are non-negative integers, and ${\cal F}_{nm}(\p_1,\ldots, \p_l)$ are functions that are independent of $\k$.  The essential feature of this expansion is that the powers of $k$ are either even or arise from contracting with the vectors $\p_i$.  This structure reflects the fact that the time evolution is local in $\delta$ and $\Phi$, and that these variables are related by $\nabla^2 \Phi = \delta$.  Although the general solution is not strictly local, any nonlocality only gives rise to even powers of $k$.  After integrating over the momenta $\p_{i}$, the factors of $(\k \cdot \p_i)^{s_i}$ in (\ref{equ:gnm}) must lead to terms that are either proportional to $(k p \mu)^{s_i}$ or vanish ($k^2$ is not possible because it would be inconsistent with $p_i \ll k$). 
  Using~(\ref{equ:ff2}), the function $f_{nm}(k)$ is then given by a sum of terms of the form 
\beq
f_{nm}(k) \supset   
 \int \frac{\d^3 p}{(2 \pi)^3}\,  2\hskip 1pt{\cal G}_{nm}(p) \, k^{2 t} \times \left\{\begin{array}{ll}  \displaystyle  + (k \hskip 1pt p\hskip 1pt \mu)^{\Delta}  \sin(\mu p r_s) \, , &  \quad \Delta= {\rm odd},\\[10pt]
-   \displaystyle   (k \hskip 1pt p\hskip 1pt \mu)^{\Delta} \,\frac{p}{k}\,  pr_s \cos(\mu p r_s) \, \frac{1-\mu^2}{2} \, ,&\quad \Delta= {\rm even},
\end{array} \right. 
\eeq
where $\Delta \leq  \sum_i s_i $ is a non-negative integer and ${\cal G}_{nm}(p)$ is independent of $k$.  We have dropped a contribution proportional to $k\hskip 1pt\partial_k \ln \Plin(k)$ that is of the same form as the $\Delta = {\rm even}$ terms.  We see that for either choice of $\Delta$, the function {\it $f_{nm}(k)$ only contains odd powers of $k$}.  Similarly, $\tilde f_{nm}(k)$ only contains even powers of $k$.
This excludes $f(k) =\sum f_{nm}(k)= const.$ as a solution, proving that a constant phase shift {\it cannot} arise.

\subsubsection{Nonperturbative Argument} 
\label{sec:NP}

The two key elements of our perturbative argument were locality (in $\Phi$) and rotational invariance.  These are features of the evolution that can be defined nonperturbatively in terms of properties of the response function $G(\p, \k-\p\hskip 2pt;\tau)$.   
This suggests that we may be able to prove that the BAO phase is protected without appealing to perturbation theory.  Such a statement is of interest because ultimately we hope to apply these results to modes that are in (or near) the nonlinear regime.  While it is true that long modes are well-described by linear evolution, one might worry that the coupling between long and short modes is no longer captured accurately by perturbation theory.  Fortunately, a soft limit consistency condition for adiabatic initial conditions~\cite{Creminelli:2013mca}  provides a nonperturbative definition of the mode coupling which will allow us to prove that the phase shift is protected even beyond perturbation theory.  

\paragraph{Causality and analyticity} Let us first remind ourselves of the link between causality in real space and analyticity in momentum space. Consider a system in which a perturbation propagates a maximal distance $R_*(\tau)$ in a time $\tau$. Causality then requires that the response function vanishes at separations greater than $R_*$, i.e.~$G({\color{Red}\x},{\color{Blue}\x-\x\hskip 1pt'}) = 0$, for $|\x-\x\hskip 1pt'| > R_*$. This implies that $G({\color{Red}\p}, {\color{Blue}\k -\p}\hskip 2pt)$ is an analytic function in the second argument ${\color{Blue}\k -\p}$.  To see this, consider
\beq
G(\x,\k-\p\hskip 2pt) = \int \d^3 y \, e^{-i(\k-\p\hskip 1pt)\cdot \y}\,G(\x,\y\hskip 1pt) \approx  \int \d^3 y  \, G(\x, \y\hskip 1pt) \big(1- i (\k-\p\hskip2pt) \cdot \y + \cdots\big) \, . \label{equ:33}
\eeq
Since the range of integration is finite, every coefficient in the series is finite.  
Provided that the series in (\ref{equ:33}) converges, the response function $G(\p,\k-\p\hskip 2pt)$ will therefore be an analytic function in $\k-\p$.

\paragraph{Local evolution without gravity}

In a time $\tau$, dark matter particles on average travel a distance $R_* \equiv v \tau$, where $v$ is the velocity dispersion of the dark matter.  Let us, for the moment, imagine an alternative reality in which only the local density of particles affects the subsequent evolution, i.e.~we ignore the long-range force from these particles.  By the argument just presented, causality guarantees that, for $|\k-\p\hskip 2pt | < R_*^{-1} \equiv k_*$, we can Taylor expand the dark matter response function 
\beq
G(\p, \k -\p\hskip 2pt)\, =\, \sum_r G_{i_1\ldots i_{r}}(\p \hskip 2pt)\,  \frac{(\k-\p\hskip 2pt)^{i_1}}{k_*} \cdots \frac{(\k -\p\hskip 2pt)^{i_r}}{k_*} \, , \label{equ:G0}
\eeq
where $G_{i_1\ldots i_{r}}(\p \hskip 2pt) \,\equiv\, (-i)^r  k_*^r\, \partial_{\q_{i_1}} \ldots \partial_{\q_{i_r}}  G(\p,\q\hskip 2pt)|_{\q = 0}$. In perturbation theory, this would simply be equivalent to the statement that the evolution is local in $\delta$; note that  (\ref{eq:gnm0})  would be of the form (\ref{equ:G0}) if $F_n F_m$ was a polynomial in all of the momenta.     
The key difference to our perturbative approach is that we will not assume a perturbative formula in terms of $\delta_{\rm in}$.   As a consequence, we cannot write our result in terms of factors of $\Plin(k)$.  Instead, the kernel in (\ref{eq:Pwlow}) takes  the form 
\begin{align} 
g(\p,\k-\p \hskip 2pt) &\,\equiv\,  \frac{\langle   G(\p, \k-\p\hskip 2pt) \,  \delta^{\rm nw}_{\rm in}(\k-\p\hskip 2pt)\, \delta^{\rm nw}(-\k\hskip 1pt)  \rangle'}{\Plin(k)} \nonumber \\[4pt] 
&\,=\, \sum_{r} \, \frac{\langle    G_{i_1 \ldots i_{r}}(\p \hskip 2pt) \,  \delta_{\rm in}^{\rm nw}(\k -\p\hskip 2pt)\, \delta^{\rm nw}(-\k\hskip 1pt)   \rangle'}{\Plin(k)}  \times  \,  \frac{(\k-\p\hskip 2pt)^{i_1}}{k_*} \cdots \frac{(\k -\p\hskip 2pt)^{i_r}}{k_*}   \, .\label{eqn:approxanalytic}  
 \end{align}
We see that the problem has reduced to computing $\langle    G_{i_1 \ldots i_{r}}(\p \hskip 2pt) \,  \delta_{\rm in}^{\rm nw}(\k -\p\hskip 2pt)\, \delta^{\rm nw}(-\k\hskip 1pt)   \rangle'$, 
where $G_{i_1 \ldots i_r}(\p\hskip 2pt)$ is independent of $\k$.
In the limit $p \ll k$, this is the squeezed limit of a bispectrum which we can simplify by using 
\begin{align}
  \delta_{\rm in}^{\rm nw}(\k -\p\hskip 2pt)\, \delta^{\rm nw}(-\k\hskip 1pt)  &= \nonumber \\[2pt]
  &\hspace{-2.5cm}= \big[  \delta_{\rm in}^{\rm nw}(\k\hskip 1pt)\,\delta^{\rm nw}(-\k\hskip 1pt) \big] \,+\, \Phi(-\p\hskip 2pt) \frac{\partial}{\partial \Phi(-\p\hskip 2pt)}\big[  \delta_{\rm in}^{\rm nw}(\k -\p\hskip 2pt)\, \delta^{\rm nw}(-\k\hskip 1pt) \big]\Big|_{\p=0} \,+\, \cdots\nonumber \\
  &\hspace{-2.5cm}\approx   \big[  \delta_{\rm in}^{\rm nw}(\k\hskip 1pt)\,\delta^{\rm nw}(-\k\hskip 1pt) \big]   \,+\,\left(-2 + \tau\hskip 1pt \partial_\tau + \k \cdot \frac{\partial}{\partial \k} \right)\,\big[  \delta_{\rm in}^{\rm nw}(\k\hskip 1pt)\, \delta^{\rm nw}(-\k\hskip 1pt) \big] \Phi(-\p\hskip 2pt)\Big|_{\p=0} + \cdots  .
  \label{equ:120}
\end{align}
In the last line, we have used the consistency condition for the adiabatic mode~\cite{Creminelli:2013mca} to replace~$\partial_\Phi$.
 We assume that the cross spectrum, $P_{\rm c}^{\rm nw}(k,\tau) \equiv \langle \delta^{\rm nw}_{\rm in}(\k\hskip 1pt) \,  \delta^{\rm nw}(-\k,\tau)\rangle'$, scales roughly as a power law in momentum, $k \,\partial_{k} \, P^{\rm nw}_{\rm c}(k) \sim n_s \,P^{\rm nw}_{\rm c}(k)$ and in time, $\tau \,\partial_{\tau} \, P^{\rm nw}_{\rm c}(k) \sim n_\tau \,P^{\rm nw}_{\rm c}(k)$, where $n_s, n_\tau = {\cal O}(1)$.    Substituting (\ref{equ:120}) into (\ref{eqn:approxanalytic}) and ignoring the constant of proportionality, we get
\beq
g(\p,\k-\p\hskip 2pt) \, \approx\, \frac{P_{\rm c}^{\rm nw}(k)}{\Plin(k)}\, \sum_{r,s}\, \left( \frac{(\k-\p\hskip 2pt) \cdot \p}{k_*p_*} \right)^{r-s} \frac{|\k-\p\hskip 2pt|^{2s}}{k_*^{2s}} \times G_{rs}(p)  \, , \label{equ:38}
\eeq
where we have introduced 
\beq
\langle G_{i_1\ldots i_r}(\p\hskip 2pt)\, \Phi(-\p\hskip 2pt) \rangle' \equiv \sum_{s=0}^{\lfloor r/2 \rfloor} \frac{\p_{i_1} \ldots \p_{i_{r-2s} }}{p_*^{r-2s}} \,\delta_{i_{r-2s +1} i_{r-2s +2}} \cdots \delta_{i_{r-1} i_{r}}\,G_{rs}(p) \, ,
\eeq
with $p_*$ being an arbitrary reference scale to make $G_{rs}(p)$ dimensionless. The sum over $s$ captures all the terms consistent with rotational symmetry. 
Notice that we have completely determined the $\k$-dependence of the function $g(\p,\k-\p\hskip 2pt) $. Since the overall factor in (\ref{equ:38}), $P^{\rm nw}_{\rm c}/\Plin$, is common to both $f(k)$ and $\tilde f(k)$ it doesn't contribute to the phase shift. We will drop it to reduce clutter. Substituting (\ref{equ:38}) into (\ref{equ:ff}), we find a sum over terms of the form
 \beq
 f(k) \supset   
  \int \frac{\d^3 p}{(2 \pi)^3}\,  2\hskip 1pt {\cal G}_{rs}(p)\, k^{2t} \times \left\{\begin{array}{ll}  \displaystyle  +\left( \frac{ k p\mu}{k_* p_*} \right)^{\Delta}  \sin(\mu p r_s) \, , &  \quad \Delta= {\rm odd},\\[10pt]
  \displaystyle  - \left( \frac{ k p\mu}{k_* p_*} \right)^{\Delta} \,\frac{p}{k}\,p r_s   \cos(\mu p r_s) \, \frac{1-\mu^2}{2} \, ,&\quad \Delta= {\rm even},
\end{array} \right. \label{equ:fresult}
\eeq
where $0 \leq \Delta \leq r$ and $0 \leq t \leq s$ are integers, and ${\cal G}_{rs}(p) \equiv (-p^2/p_*^2)^{r-t-\Delta} (p_*/k_*)^{2s}G_{rs}(p)$.  We see that $f(k)$ {\it only contains odd powers of} $k$.  A similar analysis shows that $\tilde f(k)$ in (\ref{equ:tff}) only contains even powers of $k$.  These results exclude the possibility of a constant phase shift.

\paragraph{Local evolution with gravity} 

The fact that gravity is a long-range force naively threatens to invalidate our causality argument.  Specifically, moving around dark matter particles in some region of space instantaneously changes the gravitational potential~$\Phi$ at large distances.  However, 
since physical changes to the system must conserve mass and momentum, the effect of perturbations does not propagate instantaneously. Indeed, following~\cite{Peebles:1994xt}, we will now show that any non-trivial information encoded in the initial conditions  propagates locally when expressed in terms of $\Phi$.  This is sufficient to guarantee that the phase is preserved.

 \vskip 4pt
In the Newtonian limit, we have the following nonperturbative relationship between the total density contrast and the (rescaled) gravitational potential:   
\beq
\nabla^2 \Phi = \delta \, . \label{equ:P}
\eeq
Suppose that we perturb the system at a time $\tau_0$ by changing  $\delta(\tau_0)$ to $\delta(\tau_0)+ \delta^{\rm w}(\tau_0)$ inside some region $x \leq R_*$, but without changing the total mass and the center of mass.  
Notice that, because mass and momentum are conserved, these properties are maintained at all subsequent times.  Fourier transforming $\delta^{\rm w}(\x\hskip 1pt)$ and expanding for $k R_* < 1$, we find
\beq
\delta^{\rm w}(\k\hskip 1pt) = \int \d^3 x \, e^{- i \k \cdot \x}\, \delta^{\rm w}(\x\hskip 1pt) =  \int \d^3 x \left[1-i \k \cdot \x - (\k \cdot \x\hskip 1pt)^2+ \ldots \right]  \delta^{\rm w}(\x\hskip 1pt) = {\cal O}(k^2) \, ,
\eeq
where, in the last step, we have used that $\delta^{\rm w}(\x\hskip 1pt)$ does not change the total mass and the center of mass to remove the ${\cal O}(k^0)$ and ${\cal O}(k^1)$ terms, respectively.  
Equation (\ref{equ:P}) then implies
\beq
\Phi^{\rm w}({\k}\hskip 1pt) = {\cal O}(k^0)\, . \label{eq:philocal} 
\eeq 
The absence of inverse powers of $k$ shows that the potential $\Phi^{\rm w}({\k}\hskip 1pt)$ is analytic in $\k$, for $k<R_*^{-1}$, and that the response of $\Phi^{\rm w}(\x\hskip 1pt)$ is local, i.e.~can be described as an expansion in gradients for $ x > R_*$.  

\vskip 4pt
The locality in real space can also be expressed as a restriction on the response function for $\Phi(\x\hskip 1pt)$, namely $G_{\Phi}(\x,\x-\x^{\hskip 1pt \prime}) = 0$, for $|\x-\x^{\hskip 1pt \prime}| > R_*$.  Following the previous logic, we are therefore led to the conclusion that the response function for $\Phi$ can be written as
\beq
G_\Phi(\p, \k -\p\hskip 2pt)\, =\, \sum_r G_{\Phi,i_1\ldots i_{r}}(\p \hskip 2pt)\,  \frac{(\k-\p\hskip 2pt)^{i_1}}{k_*} \cdots \frac{(\k -\p\hskip 2pt)^{i_r}}{k_*} \, . \label{equ:344}
\eeq
We note that the expansion in $\k - \p$ in (\ref{equ:344}) is equivalent to the expansion in $\k$ in~(\ref{eq:philocal}).  Moreover, since $\Phi = \nabla^{-2}  \delta$, the relationship between the response functions for $\Phi$ and $\delta$ is
\beq
G_\Phi(\p, \k -\p\hskip 2pt )\,\Phi(\k -\p\hskip 2pt)\propto G_\Phi(\p, \k -\p\hskip 2pt)\, \frac{\delta(\k -\p\hskip 2pt)}{|\k-\p\hskip 2pt|^2} = G(\p, \k -\p\hskip 2pt)\,\delta(\k -\p\hskip 2pt)\, .
\eeq 
We see that any non-analyticity in $G$ will only be in the form of additional factors of $k^{-2}$. 
Crucially, these factors do {\it not} change the odd/even counting in the powers of $k$ in (\ref{equ:ff}) and (\ref{equ:tff}), and our conclusion that the BAO phase is protected is therefore unchanged.

\subsection{Caveats}

The argument of the previous section was based on three assumptions whose violations could allow for loopholes to our result:
\begin{itemize}
\item {\it Adiabaticity.}---We used the consistency condition of the adiabatic mode~\cite{Creminelli:2013mca} to determine the coupling between long and short modes in (\ref{equ:120}).  Allowing for arbitrary initial conditions, we could, in principle, choose this mode coupling to have a non-analytic momentum dependence that would violate the even/odd counting in (\ref{equ:fresult}).  
Note that non-adiabatic initial conditions can also alter the initial phase shift~\cite{Baumann:2015rya}.   

\item {\it Growing mode.}---We implicitly assumed that the BAO phase is determined by the growing modes of the primordial fluctuations.  The growing modes for the fluctuations in the baryon and dark matter densities are the same, with initial conditions that are fixed at the time of recombination (see e.g.~\cite{Slepian:2015zra} for an analytic treatment).  For this reason, we were able to combine baryons and dark matter into the total matter overdensity with a single common phase.  On the other hand, the overall phase of a decaying mode, such as $\delta_{\rm b} - \delta_{\rm dm}$,  need not match the phase of the growing mode.  Moreover, if galaxies were biased tracers of the decaying mode, then the phase that appears in the galaxy power spectrum would be sensitive to the bias coefficient and could not be related to fundamental physics unambiguously.  Although the amplitude of the decaying mode is small, it is perhaps not negligible when extreme precision is needed. 

\item {\it Locality.}---We did not allow for dramatic nonlocality on the scale of the BAO feature and assumed that gravity is the only long-range force relevant for structure formation. Of course, if a local matter overdensity could influence the matter distribution at distances comparable to the BAO scale, one could imagine re-arranging the distribution of particles to mimic the effect of the phase shift.   
The question is whether such a re-arrangement could arise physically.  We showed that if particles move slowly and gravity is the only long-range force, then this is not possible.  On the other hand, if either a new  long-range force or the effects of propagation gave rise to $G(\p, \k -\p\hskip 2pt) \propto |\k-\p\hskip 2pt|^{-1}$, then we would get $2t \to 2t-1$ in~(\ref{equ:fresult}).  As a result, 
we would find a constant contribution to $f(k)$ [for $\Delta = 1$ and $t=0$] and hence a constant phase shift.  

In this example, the long-wavelength modes are playing a crucial role by changing the shape and peak location of the BAO feature [cf.~(\ref{equ:fresult})] and the nonlocality of $G(\p, \k -\p\hskip 2pt)$ is only modifying this change to mimic a constant phase shift.
It is natural to wonder if nonlocal evolution alone could also produce a phase shift.  
However, in the absence of the large-scale inhomogeneity provided by the long modes, the response function takes the form 
  $G(\p, \k -\p\hskip 2pt)  = (2 \pi)^3\delta_D(\p\hskip 2pt) G(\k\hskip 1pt)$ and hence $\delta^{\rm w}(\k\hskip 1pt) = A\, G(\k\hskip 1pt) \sin(k r_s )\, \delta^{\rm nw}_{\rm in}(\k\hskip 1pt)$.  The only way that $G(\k\hskip 1pt)$ could move the zeros of $\sin(k r_s)$ is if it was singular at the same points, which cannot arise physically.  We see that, without the inhomogeneity induced by the long modes, even nonlocal evolution on very large scales is insufficient to change the phase.  
\end{itemize}
These caveats are sufficiently concrete that we view them as an added opportunity to use the BAO phase as a probe of more dramatic new physics, such as non-standard initial conditions, large-scale nonlocality and new long-range forces.

\section{Conclusions} 
\label{sec:Conclusions}

In this paper, we have argued that there is an opportunity for extracting information from the spectrum of baryon acoustic oscillations that is usually discarded.
Most current BAO analysis derive their cosmological constraints from the position of the BAO peak, but not from its shape. This is natural since (after reconstruction~\cite{Eisenstein:2006nk}) the peak location is robust to the effects of gravitational nonlinearities, while the shape is distorted by a number of nonlinear effects. However, as we have shown, a part of the BAO shape contains information about the phase of the spectrum in Fourier space and this phase information does not get modified by nonlinear evolution (even without reconstruction).
 We have verified this claim in three distinct ways:
 \begin{itemize}
 \item We first showed that the phase shift corresponds to a characteristic sign change in the correlation function 
 and that 
  this feature cannot be generated by gravitational evolution.
 \item 
 We then proved that nonlinear dynamics cannot produce a phase change in the BAO spectrum to all orders in cosmological perturbation theory.
 \item Finally, we demonstrated that basic analytic properties of the linear response function allow for a nonperturbative generalization of our argument.
 \end{itemize}
Our result suggests that any physical effects in the early universe that contribute to the initial phase of the BAO spectrum can be extracted reliably at late times and will not be limited by the theoretical uncertainties that affect other large-scale structure observables~\cite{Baldauf:2016sjb}.  Weakly interacting light particles are a natural possibility for a phase shift~\cite{Bashinsky:2003tk, Baumann:2015rya}, but there may be other not yet considered physical effects. 
 
  \vskip 4pt 
As the total available information in the CMB is being saturated~\cite{Abazajian:2016yjj}, improving the sensitivity to some theoretical targets will rely on complementary data sets.   Future large-scale structure surveys have the raw statistical power to compete with the CMB and our work suggests that this can be fully exploited if the information is encoded in the phase of the BAO spectrum.  Indeed,  there is reason to be optimistic that future galaxy surveys could achieve competitive constraints on parameters like the effective number of relativistic species,~$\Neff$.  Using the galaxy and weak lensing power spectra in the linear regime, DESI and LSST should improve limits  on $\Neff$ by up to a factor of four, relative to the current Planck constraint~\cite{Font-Ribera:2013rwa}.  Moreover, future LSS surveys such as a billion object apparatus (BOA) could  exceed the limit of Planck by even a factor of 10 to 15~\cite{Dodelson:2016wal}, which would be competitive with the forecasted sensitivity of future CMB Stage 4 experiments~\cite{Abazajian:2016yjj}.  Preliminary forecasts suggest that most of this information is encoded in the BAO signal and not in the smooth  features of the (no-wiggle) power spectrum~\cite{future}.  
The promise of BAO measurements as a probe of new physics is sufficiently enticing that it warrants further investigation.

\subsubsection*{Acknowledgements}

We thank Matteo Biagetti, Raphael Flauger, Joel Meyers, Mehrdad Mirbabayi, Enrico Pajer, Rafael Porto, Uro\v s Seljak, Zachary Slepian and Benjamin Wallisch for useful discussions. 
D.B.~thanks Benjamin Wallisch for producing Figs.~\ref{fig:PhaseShift} and~\ref{fig:PS2}.
Matteo Biagetti and Benjamin Wallisch kindly provided comments on a draft.  M.Z.~is supported by the NSF grants PHY-1213563, AST-1409709, and PHY-1521097.  The work of D.B.~is part of the Delta ITP consortium, a program of the Netherlands Organisation for Scientific Research (NWO) that is funded by the Dutch Ministry of Education, Culture and Science (OCW). D.B.~also acknowledges support from a Starting Grant of the European Research Council (ERC STG Grant 279617).
\clearpage
\phantomsection
\addcontentsline{toc}{section}{References}
\bibliographystyle{utphys}
\bibliography{BAO-Refs}

\end{document}